\documentclass[11pt]{article}
\usepackage{blois,epsfig}

\usepackage{amsmath}
\usepackage{amsfonts}
\usepackage{graphicx}
\usepackage{graphicx}
\usepackage{dcolumn}
\usepackage{bm}
\usepackage[dvips]{color}
\usepackage{slashed}
\newcommand{\beq}{\begin{equation}}
\newcommand{\eeq}{\end{equation}}

\def\be{\begin{equation}}
\def\ee{\end{equation}}
\def\bea{\begin{eqnarray}}
\def\eea{\end{eqnarray}}

\begin{document}
\vspace*{4cm}
\title{The Limits of Custodial Symmetry \footnote{Speaker at conference: R. Sekhar Chivukula.
This report is a shortened version of previously published
work.\protect\cite{SekharChivukula:2009if}}}

\author{R. Sekhar Chivukula, Stefano Di Chiara, Roshan Foadi, and Elizabeth H. Simmons}
%
\address{Department of Physics,
Michigan State University, East Lansing, MI 48824, USA}
\date{\today}

\maketitle\abstracts{
We introduce a toy model implementing the proposal of using a custodial symmetry to protect the $Z b_L \bar{b}_L$ coupling from large corrections.  This ``doublet-extended standard model" adds a weak doublet of fermions (including a heavy partner of the top quark) to the particle content of the standard model in order to implement an $O(4) \times U(1)_X  \sim SU(2)_L \times SU(2)_R \times P_{LR} \times U(1)_X$ symmetry that protects the $Z b_L \bar{b}_L$ coupling.  This symmetry is softly broken to the gauged $SU(2)_L \times U(1)_Y$ electroweak symmetry by a Dirac mass $M$ for the new doublet; adjusting the value of $M$ allows us to explore the range of possibilities between the $O(4)$-symmetric ($M \to 0$) and standard-model-like ($M \to \infty$) limits.}

\section{Introduction}
Agashe \cite{Agashe:2006at} et al.  have shown that the constraints on beyond the standard model physics related to the $Zb_L\bar{b}_L$ coupling can, in principle, be loosened if the global $SU(2)_L \times SU(2)_R$ symmetry of the electroweak symmetry breaking sector is actually a subgroup of a larger global symmetry  of both the symmetry breaking and top quark mass generating sectors of the theory. In particular, they propose that these interactions preserve an  $O(4)\sim SU(2)_L\times SU(2)_R\times P_{LR}$ symmetry, where $P_{LR}$ is a parity interchanging $L\leftrightarrow R$. The $O(4)$ symmetry is then spontaneously broken to  $O(3)\sim SU(2)_V\times P_{LR}$,  breaking the elecroweak interactions but
protecting $g_{Lb}$ from radiative corrections, so long as the left-handed bottom quark is a $P_{LR}$ eigenstate.

In this talk we report on the construction of the simplest $O(4)$-symmetric extension of the SM.\cite{SekharChivukula:2009if}, the doublet-extended standard model or DESM.  

\subsection{The Model}

We extend the global $SU(2)_L \times SU(2)_R$ symmetry of the Higgs sector of the SM to an $O(4)\times U(1)_X\sim SU(2)_L\times SU(2)_R\times P_{LR} \times U(1)_X$ for both the symmetry breaking and top quark mass generating sectors of the theory.  As usual, only the electroweak subgroup, $SU(2)_L\times U(1)_Y$, of this global symmetry is gauged; our model does not include additional electroweak gauge bosons.  The global $O(4)$ spontaneously breaks to $O(3) \sim SU(2)_V \times P_{LR}$ which will protect $g_{Lb}$ from radiative corrections,\cite{Agashe:2006at}  provided that the left-handed bottom quark is a parity eigenstate:  $P_{LR} b_L = \pm b_L$.  The additional global $U(1)_X$ group is included to ensure that the light $t$ and $b$ eigenstates, the ordinary top and bottom quarks, obtain the correct hypercharges.

We therefore introduce a new doublet of fermions $\Psi  \equiv (\Omega, T^\prime)$.  The left-handed component, $\Psi_L$ joins with the top-bottom doublet $q_L \equiv (t_L^\prime, b_L)$ to form an $O(4)\times U(1)_X$ multiplet
\begin{equation}
  {\cal Q}_L = \left( {\begin{array}{*{20}c}
   {t^\prime_L } & {\Omega_L }  \\
   {b_L } & {T^\prime_L }  \\
 \end{array} } \right)
\equiv \left(\begin{array}{cc} q_L & \Psi_L \end{array}\right) \ ,
\end{equation}
which transforms as a $(2,2^*)_{2/3}$ under $SU(2)_L\times SU(2)_R\times U(1)_X$.  The parity operation $P_{LR}$, which exchanges the $SU(2)_L$ and $SU(2)_R$
transformation properties of the fields, acts on ${\cal Q}_L$ as:
\begin{equation}
P_{LR} {\cal Q}_L = - \left[ ( i \sigma_2)\, {\cal Q}_L\, (i \sigma_2) \right]^T = \left( {\begin{array}{*{20}c}
   {T^\prime_L } & {-\Omega_L }  \\
   {-b_L } & {t^\prime_L }  \\
 \end{array} } \right)
\end{equation}
exchanging the diagonal components, while reversing the signs of the off-diagonal components. The $t^\prime$ and $T^\prime$ states mix to form mass eigenstates corresponding to the top quark ($t$) and a heavy partner ($T$).  

\begin{table}[bt]
\caption{\label{tab:charges} Charges of the fermions under the various symmetry groups in the model. Note that, as discussed in the text, other $T^3_R$ and $Q_X$ assignments for the $\Omega_R$ and $T^\prime_R$ states are possible.}
\begin{center}
\begin{tabular}{|c||c|c|c|c||c|c|c|c|}
\hline
&$t^\prime_L$&$b_L$&$\Omega_L$ & $T^\prime_L$&$t^\prime_R$&$b_R$&$\Omega_R$ & $T^\prime_R$\\
\hline
$T^3_L$& $\frac12$ & $-\frac12$ & $\frac12$ & $-\frac12$ & $0$ & $0$ & $\frac12$ & $-\frac12$  \\
\hline
$T^3_R$& $-\frac12$ & $-\frac12$ & $\frac12$ & $\frac12$ & $0$ & $-1$ & $0 $ & $0$\\
\hline
$Q$& $\frac23$ & $-\frac13$ & $\frac53$ &$\frac23$ &$\frac23$ & $-\frac13$ & $\frac53$ &$\frac23$ \\
\hline
$Y$& $\frac16$ & $\frac16$ & $\frac76$ & $\frac76$ & $\frac23$ & $-\frac13$ & $\frac76$ &$\frac76$\\
\hline
$Q_X$& $\frac23$ &$\frac23$ &$\frac23$ &$\frac23$ &$\frac23$ &$\frac23$ &$\frac76$ &$\frac76$ \\
\hline
\end{tabular}
\end{center}
\end{table}

We assign the minimal right-handed fermions charges that accord with the symmetry-breaking pattern we envision:  the top and bottom quarks will receive mass via Yukawa terms that respect the full $O(4) \times U(1)_X$ symmetry, while the exotic states will have a dimension-three mass term that explicitly breaks the large symmetry to $SU(2)_L \times U(1)$. The charges of all the fermions are listed in Table \ref{tab:charges}.

Now, let us describe the symmetry-breaking pattern and fermion mass terms explicitly.  
Spontaneous electroweak symmetry breaking proceeds through a Higgs multiplet that transforms as a $(2,2^*)_0$ under $SU(2)_L \times SU(2)_R \times U(1)_X$:
\begin{equation}
 \Phi  = \frac{1}{\sqrt{2}}\left( {\begin{array}{*{20}c}
   {v+h + i \phi^0 } & {i\sqrt{2}\ \phi^+  }  \\
   {i\sqrt{2}\ \phi^-  } & {v+h-i\phi^{0} }  \\
 \end{array} } \right)\ .
 \label{eq:higgsdef}
\end{equation}
Again, the parity operator $P_{LR}$ exchanges the diagonal fields and reverses the signs of the off-diagonal elements.  When the Higgs acquires a vacuum expectation value, the longitudinal  $W$ and $Z$ bosons acquire mass and a single Higgs boson remains in the low-energy spectrum.
The Higgs multiplet has an $O(4)\times U(1)_X$ symmetric Yukawa interaction with the top quark:
\begin{equation}
{\cal{L}}_{\rm Yukawa}= - \lambda_t \text{Tr} \left( \overline{\cal Q}_L\cdot \Phi\right) t^\prime_R \ + \rm{h.c.}\ .
\label{eq:Yuk}
\end{equation}
that contributes to generating a top quark mass.\footnote{Here we neglect $m_b$ and any other Yukawa 
interactions.\protect\cite{SekharChivukula:2009if}}

Next we break the  full $O(4)\times U(1)_X$ symmetry to its electroweak subgroup. We do so
first by gauging $SU(2)_L\times U(1)_Y$.  In addition, we wish to preserve the $O(4)$ symmetry
of the top quark mass generating sector in all dimension-4 terms, but break it 
softly by introducing a dimension-3 Dirac mass term for  $\Psi$,
\begin{equation}
{\cal{L}}_{\rm mass}= - M\ \bar{\Psi}_L\cdot\Psi_R + h.c. 
\label{eq:M}
\end{equation}
that explicitly breaks the global symmetry to $SU(2)_L\times U(1)_Y$.  Note that we therefore expect that any flavor-dependent radiative corrections to the $Zb_L\bar{b}_L$ coupling will vanish in the limit $M \to 0$, as the protective parity symmetry is restored; alternatively, as $M \to \infty$, the larger symmetry is pushed off to such high energies that the resulting theory looks more and more like the SM.

\subsection{Mass Matrices and Eigenstates}

When the Higgs multiplet acquires a vacuum expectation value and breaks the electroweak symmetry, masses are generated for the top quark, its heavy partner $T$ and the exotic fermion $\Omega$ through the mass matrix:
\begin{equation}
{\cal L}_{\rm mass} = -\left(\begin{array}{*{20}c} t^\prime_L & T^\prime_L \end{array}\right)\ 
\left({\begin{array}{*{20}c} m & 0  \\ m & M \\ \end{array} } \right)
\left(\begin{array}{c} t^\prime_R \\ T^\prime_R \end{array}\right) -M \bar{\Omega}_L \Omega_R + {\rm h.c} \ ,
\end{equation} 
where
\begin{equation}
m = \frac{{\lambda _t v}}{{\sqrt 2 }} \ .
\label{eq:mtSM}
\end{equation} 

Diagonalizing the top quark mass matrix yields mass eigenstates $t$ (corresponding to the SM top quark) and $T$ (a heavy partner quark), with corresponding eigenvalues
\begin{equation}
m_t^2 = \frac{1}{2}\left[1-\sqrt{1+\frac{4m^4}{M^4}}\right]M^2+m^2 \ , \ \ \ \ \ \ \ \ 
m_T^2 = \frac{1}{2}\left[1+\sqrt{1+\frac{4m^4}{M^4}}\right]M^2+m^2 \ .
\label{eq:massEv}
\end{equation}
The mass eigenstates are related to the original gauge eigenstates through the rotations
whose mixing angles are given by 
\begin{equation}
\sin\theta_R=\frac{1}{\sqrt{2}}\sqrt{1-\frac{1-2m^2/M^2}{\sqrt{1+4m^4/M^4}}} \ ,\ \
\sin\theta_L=\frac{1}{\sqrt{2}}\sqrt{1-\frac{1}{\sqrt{1+4m^4/M^4}}} \ .
\end{equation}
From these equations the decoupling limit $M\to\infty$ is evident: $m_t$ approaches its SM value as in Eq.~(\ref{eq:mtSM}), the $t-T$ mixing goes to zero, and $T$ becomes degenerate with $\Omega$. Conversely, in the limit $M\to 0$, the full $O(4)\times U(1)_X$ symmetry is restored and only the combination $T'_L + t'_L$ couples to $t_R$ with mass $m$. For phenomenological discussion, it will be convenient to fix $m_t$ at its experimental value and express the other masses in terms of $m_t$ and the ratio $\mu\equiv M/m$. 

\section{$\delta g_{Lb}$, $\alpha S$, and $\alpha T$}

\begin{figure}
\caption{The solid (blue) curve shows the DESM model's prediction for $g_{Lb}$ The thick horizontal line corresponds to $g_{Lb}^{ex}=-0.4182$, while the two horizontal upper and lower solid lines bordering the shaded band correspond to the $\pm 1\sigma$ deviations \protect\cite{:2005ema}. The SM prediction is given by the dashed horizontal line. The leading-log contribution is shown by the dotted curve.}
\begin{center}
\includegraphics[width=3 in]{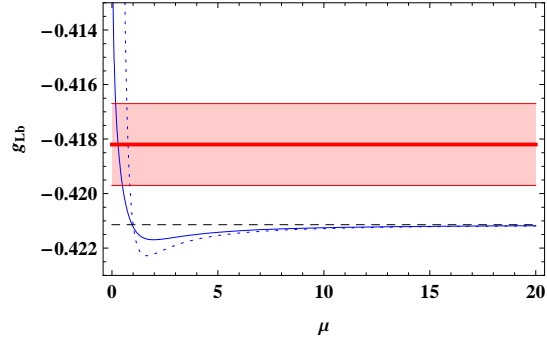}
\end{center}
\label{fig:gZbb}
\end{figure}

We now display the value of the $Z\bar{b}_L b$ coupling,  $g_{Lb}$,  in our model \cite{SekharChivukula:2009if}  (as a function of $\mu$ for fixed $m_t$), and compare with the values given by experiment and the SM, as illustrated in Fig.~(\ref{fig:gZbb}).  The (solid blue) curve shows how $g_{Lb}$ varies with $\mu$ in our model; we required $g_{Lb}$ to match the SM value with $m_t = 172$ GeV and $v = 246$ GeV as $\mu \to \infty$.   We see that $g_{Lb}$ in our model is slightly more negative than (i.e. slightly farther from the experimental value than) the SM value for $\mu > 1$, agrees with the SM value for $\mu = 1$, and comes within $\pm 1 \sigma$ of the experimental value only for $\mu < 1$.  Given the shortcomings of the small-$\mu$ limit, this is disappointing.

\begin{figure}
\caption{The dots represent the theoretical predictions of the DESM  (with $m_h$ set to the reference value $115$ GeV), showing how the values of $\alpha S$ and $\alpha T$ change as $\mu$ successively takes on the values $3,\,4,\, 5, \,...,\,20,\,\infty$.   The three ellipses enclose the $95\%$CL regions of the $\alpha S$ - $\alpha T$ plane for the fit to the experimental data performed in \protect\cite{Amsler:2008zzb}; they correspond to Higgs boson mass values of $m_h = 115\, {\rm GeV},\, 300\, {\rm GeV,\, and}\ 1\, {\rm TeV}$.   Comparing the theoretical curve with the ellipses shows that the minimum allowed value of $\mu$ is 20, for $m_h=117$ GeV. }
\begin{center}
\includegraphics[width=3 in]{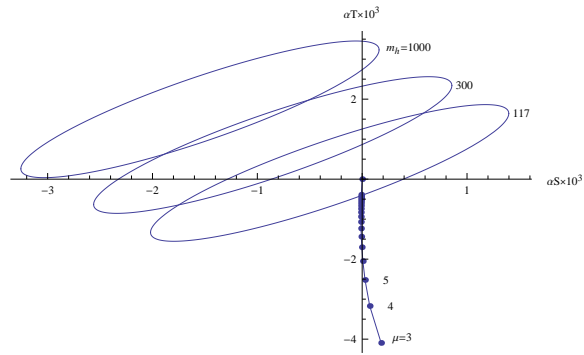}
\end{center}
\label{fig:Ellipses}
\end{figure}

Furthermore, in Figure \ref{fig:Ellipses} we show the DESM predictions\cite{SekharChivukula:2009if} for  the
oblique parameters \cite{Peskin:1991sw,Altarelli:1990zd,Altarelli:1991fk} $[\alpha S^{th}(\mu), \alpha T^{th}(\mu)]$ using $m_h = 117$ GeV,  and illustrating the successive mass-ratio values $\mu = 3,\, 4,\,...,20,\,\infty$; the point $\mu = \infty$ corresponds to the SM limit of the DESM and therefore lies at the origin of the $\alpha S$ - $\alpha T$ plane. 
From this figure, we observe directly that the $95\%$CL lower limit on $\mu$ for $m_h=115\text{ GeV}$ is about 20, while for any larger value of $m_h$ the DESM with $\mu\leq 20$ is excluded at 95\%CL. In other words, the fact that a heavier $m_h$ tends to worsen the fit of even the SM ($\mu \to \infty$) to the electroweak data is exacerbated by the new physics contributions within the DESM.
The bound $\mu\geq 20$ corresponding to a DESM with a 115 GeV Higgs boson also implies, at 95\%CL, that $m_{T}\geq \mu \ m_t \cong 3.4\text{ TeV}$, so that the heavy partners of the top quark would likely be too heavy for detection at LHC.  

\end{document}